# High-quality all-oxide Schottky junctions fabricated on heavily Nb-doped SrTiO$_3$ substrates

A. Ruotolo[*], C.Y. Lam, W.F. Cheng, K.H. Wong and C.W. Leung

*Department of Applied Physics and Materials Research Centre, The Hong Kong Polytechnic University, Hung Hom, Kowloon, Hong Kong, China*

## Abstract

We present a detailed investigation of the electrical properties of epitaxial La$_{0.7}$Sr$_{0.3}$MnO$_3$/SrTi$_{0.98}$Nb$_{0.02}$O$_3$ Schottky junctions. A fabrication process that allows reduction of the junction dimensions to current electronic device size has been employed. A heavily doped semiconductor has been used as a substrate in order to suppress its series resistance. We show that, unlike standard semiconductors, high-quality oxide-based Schottky junctions maintain a highly rectifying behavior for doping concentration of the semiconductor larger than $10^{20}$ cm$^{-3}$. Moreover, the junctions show hysteretic current-voltage characteristics.

* Corresponding author:
A. Ruotolo
Dept. of Applied Physics
Hong Kong Polytechnic University,
Hung Hom, Kowloon
Tel.:    +852 2766 5671
Fax:    +852 2333 7629
E-mail:   antonio.ruotolo@mac.com

## 1. Introduction

Perovskite-oxide Schottky junctions have recently attracted much interest because of their potential as non-volatile memory cells. These systems show electro-resistive (ER) switching that is believed to arise from the change in Schottky barrier height and/or width by trapped charge carriers at the interface states[1,2]. Unlike ER switching observed in many oxide films, in which an electric field generates conductive filamentary paths[3,4], the interface-type switching is expected to be strictly confined at the junction interface. As a consequence, the latter is more promising for high-density integration.

So far, Schottky-like systems reported in literature have been patterned in large mesa (from $100 \times 100$ μm$^2$ to several mm$^2$) with electrical contact to the top layer provided by direct bonding within the junction area. Given the high resistivity of the oxide semiconductors employed, the use of such large junctions can easily lead to a misinterpretation of the experimental data when large magnetic fields are applied. The large size of the junctions employed in experiments casts serious doubts on possible magnetoresistive effects in these systems, too often naively reported[5,6] and then refuted[7]. Current density disuniformity induced by large magnetic fields is likely to be the origin of the detected change of magnetoresistance. A fabrication process that allows the definition of much smaller junctions, each of which provided with connection wiring and contacting pads, is required to test the potentials of these systems for applications.

Another problem that must be overcome towards the launch of oxide-electronic devices is represented by the resistive contribution of the active substrate. In order to obtain highly rectifying Schottky junctions, the doping level of the active substrate is kept below $10^{19}$ cm$^{-3}$, as in standard semiconductors[10]. Unfortunately, semiconductor oxides have much larger resistivity as compared to standard semiconductors at the same doping level. As a result, the resistance offered by the substrate can be comparable to, or even larger than, that of the junction in forward bias, resulting in an overall drop of the performance of the device.

In the present study, we report on the fabrication and characterization of high quality epitaxial La$_{0.7}$Sr$_{0.3}$MnO$_3$/SrTi$_{0.98}$Nb$_{0.02}$O$_3$ (LSMO/NSTO) Schottky junctions with size ranging from 5 μm$^2$ to 9 μm$^2$. A detailed analysis of the interface transport properties as a function of temperature is presented. We show that highly rectifying Schottky junctions can be fabricated by starting from heavily doped *n*-type oxide substrates. The resistance offered by the degenerate substrate is, in this case, small enough not to affect the performance of the devices. We also show that, in high quality junctions, the ER switching do not disappear for doping level as high as that used here, in disagreement with what observed in similar systems[8] of larger size.

## 2. Fabrication process

The NSTO substrate was covered with 50 nm thick insulating epitaxial $SrTiO_3$ (STO) deposited by Pulsed Laser Deposition (PLD) except for narrow channels, that were preserved uncovered by resorting to shadowing with Copper wires (see Fig. 1a). The pulse frequency was 5 Hz and the laser fluency at the target surface was ∼ 5 J/cm$^2$. The substrate temperature was maintained at 650 °C throughout the deposition and the oxygen pressure was 150 mTorr. The film was post-annealed for 10 min under the same conditions. Wires of 70 μm in diameter could preserve channels of 4 μm in width uncovered (Fig. 1b) if a complete contact between the wire and the substrate was achieved. A large area on the NSTO substrate was also preserved for electrical contact by resorting to a metallic shadow mask. A 100 nm epitaxial LSMO layer was then grown by PLD after removing the shadow wires (Fig. 1c). Deposition and post-annealing conditions were the same as for the STO film. The LSMO layer was covered with a 50 nm thick polycrystalline Platinum (Pt) layer to provide ohmic contact. Standard UV-lithography and Argon ion milling were used to define the junctions in a cross-type configuration (Fig. 1d). Connection wires and contact pads are isolated from the active substrate by the insulating STO. Epitaxial STO has been proved[9] to have a leakage current smaller than $1 \times 10^{-6}$ A/cm$^2$ for electric field up to 1 MV/cm.

The current-voltage characteristics could be measured in our devices in both two-probe and four-probe configurations. Yet, as demonstrated in the following, the Pt layer was capable to short the additional resistance due to the LSMO wiring. Additional resistance due to the substrate is negligible because of the high level of its donor carrier concentration, $N_{NSTO}$ ∼ 0.02 [1/u.c.] = $3.3 \times 10^{20}$ cm$^{-3}$, corresponding to a measured resistivity of 3.5 mΩcm at room temperature.

## 3. Results and discussion

Current density-voltage (*J-V*) and capacitance-voltage (*C-V*) characteristics were recorded by a two-probe method in a range of temperatures $T$ = 20 - 400 K. Fig. 2 shows the *J-V* characteristics of a 1.5 × 4 µm$^2$ junction at various *T*. The junctions show a very highly rectifying behavior and a hysteretic regime (Fig. 3) sets in when the forward bias current ($J_F$) is increased up to 50 kA/cm$^2$. The hysteresis is stable over repeated sweeps. In Fig. 2, the change of voltage with respect to the temperature (d$V$/d$T$) is linear and equals to d$V$/d$T$ = 0.1 mV/K, independently of the forward bias current. This very low sensitivity to the temperature is in agreement with the behavior expected for a Schottky diode[10].

The *J-V*s were insensitive to magnetic fields up to 0.7 T, as expected from an oxygen stoichiometric LSMO/NSTO junction[7].

In a Schottky diode model, $J_F$ can be expressed as[10]:

$$J_F = J_S \exp\left(\frac{qV}{nk_BT}\right), \tag{1}$$

$$J_S = A^*T^2 \exp\left(-\frac{q\phi_B}{k_BT}\right), \tag{2}$$

where $q$ is the electron charge, $k_B$ the Boltzman's constant, $T$ the temperature, $A^*$ the Richardson's constant, $\phi_B$ the barrier height and $n$ is an ideality factor (equal to unity in a purely thermoionic emission regime). Eq. 1 is a good approximation under the restriction $V \gg k_BT/q$. The Richardson's constant will be taken as $A^* = 4\pi m^* k_B/h^3$ = 156 A cm$^{-2}$ K$^{-2}$, corresponding to an effective mass of the electrons for the NSTO[11] $m^* = 1.3\ m_0$ with $m_0$ being free electron mass. The (ln$J_F$)/$V$ characteristics (see inset Fig. 2) show only a slight bending, for large current densities, from the theoretical straight lines expected from Eq. 1. The current $J_F$ is far from saturating, hence indicating a negligible ohmic series resistance. From the slope of the straight lines, $n$ can be deduced as $n = (q/k_BT)(\partial V/\partial \ln J_F)$, and from the intercept of ln$J_F$ at $V = 0$, $\phi_B$ can be extracted. The calculated values are plotted in Fig. 4. The ideality factor indicates a significant increase of the tunneling contribution when reducing the temperature. This tunneling contribution is significant in all the range of temperatures investigated ($n = 1.73$ at $T = 400K$). The changing of the ratio between tunneling and thermoionic current components is best shown by the reverse current density ($J_R$) as a function of $V$ (magnified in Fig. 5). With the tunnel current component becoming dominant, the $J_R/V$ becomes sharper, resulting in a non-monotonic voltage at a fixed reversed bias as a function of $T$ (inset Fig. 5).

The barrier height in Fig. 4 decreases to an unphysically small value, indicating a losing of the validity of the model expressed by Eq. 1. Since electron tunneling takes place, the barrier height as calculated by Eq. 1 is physically meaningless.

The behavior of a Schottky junction when tunneling sets in has been widely studied[12,13]. At very low temperatures and for sufficiently thin barriers (i.e. highly doped semiconductors), electrons can tunnel directly from the conduction band of the semiconductor into to the metal (direct tunneling). The forward bias current density in the direct tunneling regime ($J_F^t$) as a function of $V$ can be expressed as[12]:

$$J_F^t = J_S^t \exp\left(\frac{qV}{E_{00}}\right), \qquad (3)$$

$$E_{00} = \frac{qh}{4\pi}\left(\frac{N_{NSTO}}{m^* \varepsilon_{NSTO}}\right)^{1/2}, \qquad (4)$$

where $h$ is Planck's constant, $\varepsilon_{NSTO}$ is the permittivity of the semiconductor and $J_S^t$ is only slightly dependent on the temperature and can be considered constant.

At intermediate temperatures, electrons are thermally excited and tunneling occurs around an energy level above the Fermi level (thermally assisted tunneling). The situation is depicted in Fig. 6.

The forward bias current density in the thermally assisted tunneling regime ($J_F^{th}$) as a function of $V$ can be expressed as[12]:

$$J_F^{th} = J_S^{th} \exp\left(\frac{qV}{E_0}\right), \qquad (5)$$

$$E_0 = E_{00} \coth(E_{00}/k_B T), \qquad (6)$$

$$J_S^{th} = \frac{A^* T^2 \pi^{1/2} E_{00}^{1/2}\left[q(\phi_B - V) + \xi\right]^{1/2}}{k_B T \cosh(E_{00}/k_B T)} \exp\left(\frac{\xi}{k_B T} - \frac{q\phi_B + \xi}{E_0}\right), \qquad (7)$$

where $\xi$ is the energy difference between $E_F$ and the conduction band bottom of the semiconductor. By comparing Eq. 5 with Eq. 1, the values of $E_0$ as a function of $T$ can be simply calculated from those of $n$ as extrapolated from the $J$-$V$ curves by assuming valid the model expressed by Eq. 1.:

$$\left.\frac{\partial \ln J_F^{th}}{\partial V}\right|_T = \left.\frac{\partial \ln J_F}{\partial V}\right|_T \Rightarrow E_0 = nk_BT \qquad (8)$$

The calculated values are shown in Fig. 7. The saturation limit can be estimated to be $E_{00} = 54.8$ meV by fitting the experimental data with Eq. 6.

If $\xi$ is much smaller than $q\phi_B$, from Eq. 7, the slope of $\ln\left[J_F^{th}\cosh(E_{00}/k_BT)/T\right]$ vs $1/E_0$ gives $-q\phi_B$. Eq. 7 does not hold when the devices approach direct tunneling regime[Error! Bookmark not defined.]. Besides, in heavily doped semiconductors, the condition $\xi \ll q\phi_B$, is hardly fulfilled at any temperature. For instance, in our devices, assuming a carrier concentration independent of the temperature, $\xi = (h^2/8m^*)*(3N_{NSTO}/\pi)^{3/2} = 0.13$ eV, that is of the same order as the expected barrier height. An estimation of $\phi_B$ must be obtained by measuring the capacitance in reverse bias as a function of the voltage.

Fig. 8 shows $1/C^2$-$V$ characteristics recorded at various $T$. In a Schottky diode a linear relation is expected[10]:

$$\frac{1}{C^2} = 2\frac{V_{B0} - V}{q\varepsilon_{NSTO}N_{NSTO}} \qquad (9)$$

where $C$ is the capacitance per unit area and $V_{B0}$ is the built-in potential at zero bias. At the highest temperatures, a slight deviation of the characteristics in Fig. 8 from the linear behavior is ascribable to the increase of the leakage current in reverse bias with a consequent losing of reliability of the measurement when increasing the reverse bias voltage. From the intercept of the straight lines with the $V$ axis, a built-in potential of $V_{B0} = 0.83 \pm 0.05$ V can be estimated. The intrinsic barrier height (neglecting the rounding off due to image force) can be estimated as[13] $\phi_{B0} = V_{B0} + \xi/q = 0.96$ eV. $\phi_{B0}$ is not significantly dependent on $T$, confirming that the decrease of $\phi_B$ with $T$ in Fig. 4 reflects a change in the energy of the thermal assistance, not properly considered by Eq.1, rather than a change in the barrier height.

From the slope of the straight lines the product $\varepsilon_{NSTO}(T)N_{NSTO}(T)$ can be extracted. If $N_{NSTO}$ can be considered constant with $T$ and equals to the nominal carrier concentration, $\varepsilon_{NSTO}(T)$ can be deduced as plotted in Fig 9, normalized to the vacuum permittivity $\varepsilon_0$.

$\varepsilon_{NSTO}$ is much smaller than the permittivity of undoped STO, in all the range of temperatures investigated, as well as much less temperature dependent. This is well-known[14,15] to be due to the electric field dependence of $\varepsilon_{NSTO}$ in the depletion layer of NSTO-based Schottky junctions, and therefore, will not be discussed here. What instead we want to point out is that this reduction of permittivity is not large enough to drive the junction to behave as an ohmic contact (direct tunneling regime at all temperatures), even for doping level of NSTO as large as that of the LSMO.

LSMO is a heavily doped $p$-type semiconductor that, below about room temperature, can be regarded as a metal with a carrier concentration $N_{LSMO} \sim 0.1$ [1/u.c.] = $1.65 \times 10^{21}$ cm$^{-3}$. At the interface with NSTO, a $p/n$ junction is formed with a built-in depletion layer that extends in the LSMO(NSTO) layer over a length:

$$W_{LSMO(NSTO)} = \sqrt{\frac{2\varepsilon_{LSMO(NSTO)}V_{b0}}{qN_{LSMO(NSTO)}}} \quad (10)$$

The junction has the feature of a one side abrupt $p^{++}/n$ junction, *i.e.* a Schottky junction, as long as the condition:

$$W_{LSMO} \ll W_{NSTO} \Leftrightarrow \frac{\varepsilon_{LSMO}}{N_{LSMO}} \ll \frac{\varepsilon_{NSTO}}{N_{NSTO}} \quad (11)$$

is satisfied. Although for $N_{NSTO} = 3.3 \times 10^{20}$ cm$^{-3}$, $\varepsilon_{NSTO}$ becomes of the order of several tens of $\varepsilon_0$, the permittivity of the LSMO is smaller than $10\varepsilon_0$ even at room temperature[16], and is likely to become smaller near the interface because of the presence of a large electric field. Therefore, condition (11) remains well satisfied in a large range of $T$, even for doping levels of the NSTO as large as the one used in this work.

Finally, let us observe that, from Fig. 9, the permittivity $\varepsilon_{NSTO}$ in the limit $T = 0$ K must be $\sim 30\varepsilon_0$. This implies, through Eq. 4, that at $T = 0$ K the effective carrier concentration is $\sim 3.3 \times 10^{20}$ cm$^{-3}$, that confirms the assumption made of an effective carrier concentration equals to the nominal one, in all the range of temperatures investigated, as expected for a heavily doped semiconductor. This also excludes that the formation of the Schottky barrier is due to an effectively smaller value of the carrier concentration in the NSTO close to the interface due to oxygen deficiency.

## 4. Conclusions

In conclusion, we have fabricated small high-quality epitaxial Schottky diodes of a heavily doped *p*-type semiconductor (LSMO) on a heavily doped *n*-type substrate (NSTO). Electrical transport across the interface is dominated by thermally assisted tunneling for temperatures up to 400 K. Despite the heavy doping of both the electrodes, the devices maintain a highly rectifying *J-V* characteristic in a large range of temperatures. This is ascribed to the large permittivity of the NSTO substrate as compared to that of the LSMO film. The permittivity of the NSTO remains one order of magnitude larger than that of the LSMO, even for very heavy doping concentrations of the substrate.

While removing the series resistive contribution of the substrate, our devices maintain a reproducible hysteresis in the *J-V* characteristics, that can be exploited for non-volatile memory applications.

## 5. Acknowledgments

This work was supported by the Hong Kong Polytechnic University through the grant A-PG91.

**Captions:**

**Fig. 1**: Schematic of the fabrication process: (a) insulating STO is deposited on the NSTO substrate; (b) 4μm wide channels are opened in the insulating STO by resorting to shadowing with Copper wires; (c) LSMO is deposited and covered with Pt; (d) junctions and wiring are defined by standard UV photolithography and Argon ion milling.

**Fig. 2**: *J-V* characteristics at various *T*. The inset shows the same characteristics in forward bias on a logarithmic scale.

**Fig. 3**: Hysteretic *J-V* characteristic recorded at room temperature. Three consecutive sweeps are shown. The forward bias current ($J_F$) and the absolute value of the reverse bias current ($J_R$) are reported on two different scales for the sake of clearness.

**Fig. 4**: Ideality factor *n* and barrier height $\phi_B$ as calculated by Eq. 1. The lines are guides to the eye.

**Fig. 5**: *J-V* characteristics in reverse bias at various *T*. The inset shows the voltage at a fixed bias $J_{bias} = -0.2$ kA/cm$^2$ as a function of *T*.

**Fig. 6**: Schematic potential profile for the Schottky junctions in forward bias.

**Fig. 7**: Energy $E_0$ as a function of *T*. A value of $E_{00} = E_0(0) = 54.8$ meV is estimated by fitting the data with Eq. 6.

**Fig. 8**: $1/C^2$-*V* characteristics measured at a frequency of 1 kHz.

**Fig. 9**: Permittivity of the NSTO in the depletion layer, as estimated from the slope of the $1/C^2$-*V* characteristics.

Fig. 1

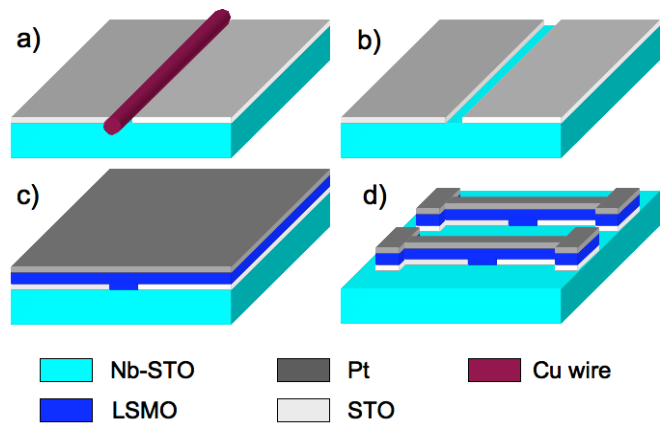

Fig. 2

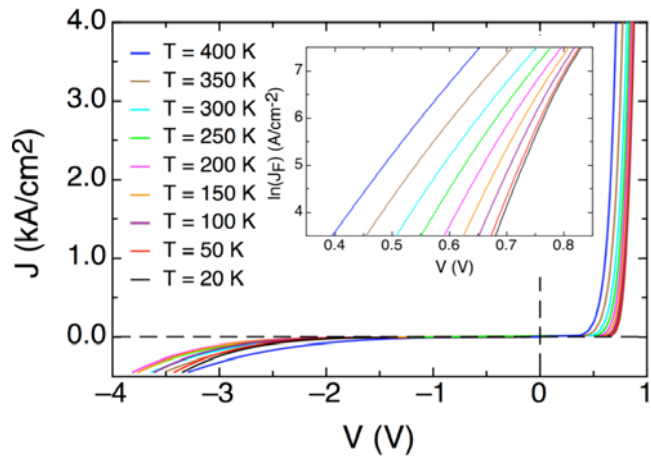

Fig. 3

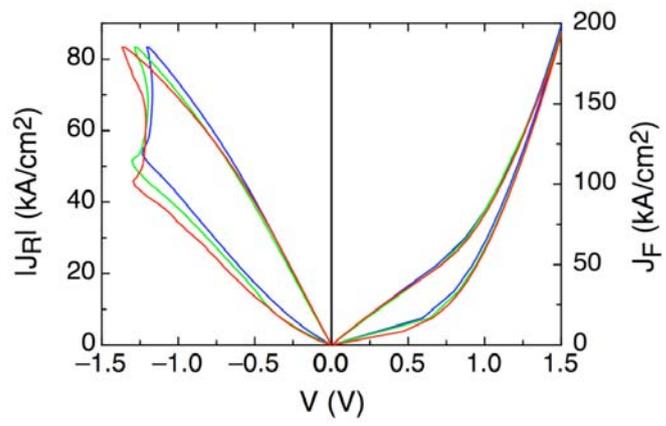

Fig. 4

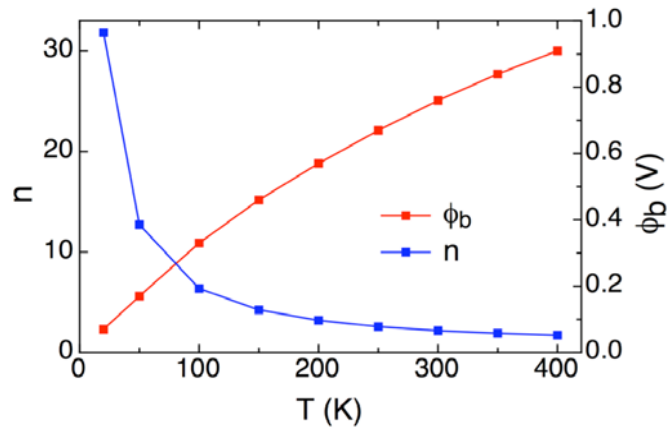

Fig. 5

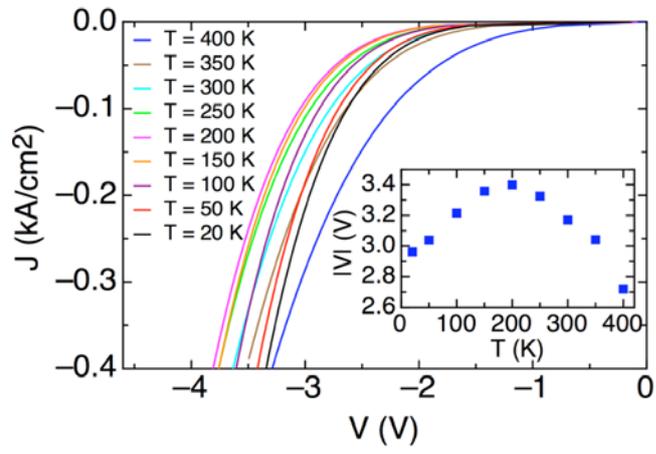

Fig. 6

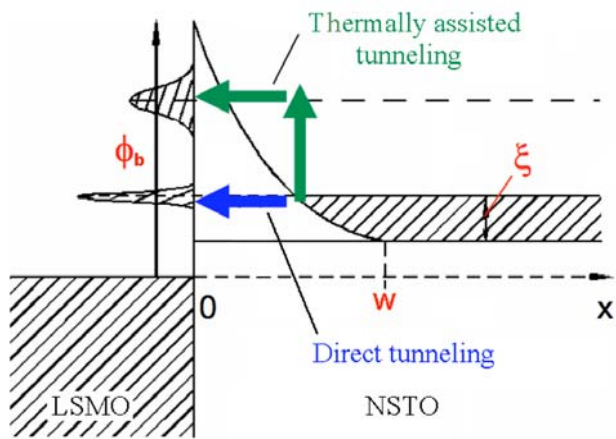

Fig. 7

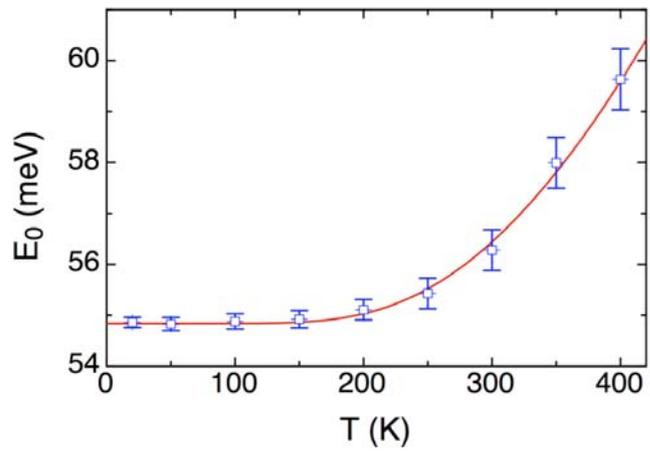

Fig. 8

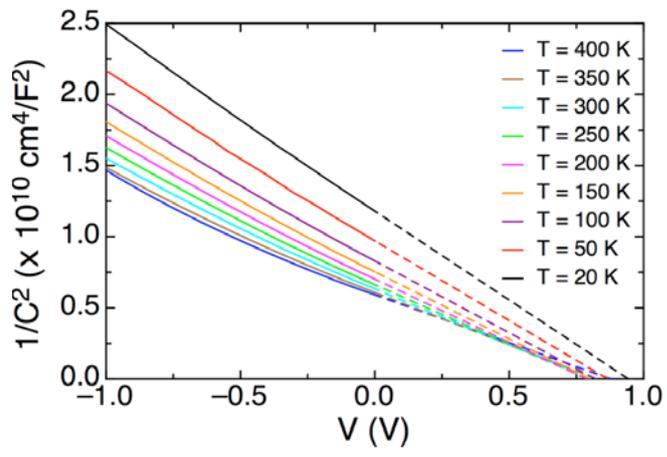

Fig. 9

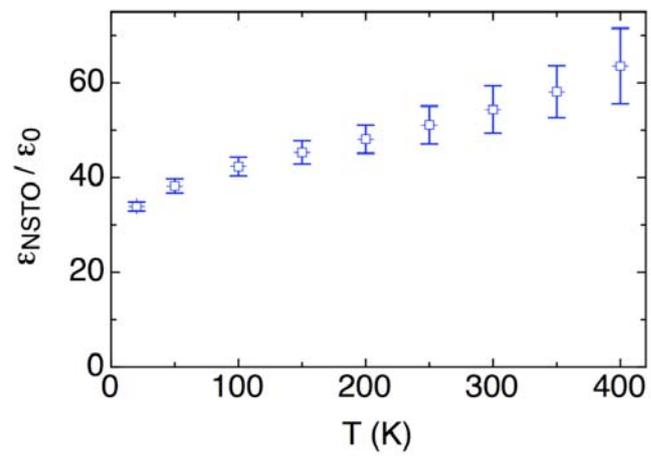